\documentclass[11pt,american,twoside,a4paper]{article}
\usepackage[T1]{fontenc}
\usepackage[utf8]{inputenc}
\usepackage{latexsym}
\usepackage{amsmath}
\usepackage{bbm}
\usepackage{amssymb}
\usepackage{babel}
\usepackage{amsfonts}
\usepackage{fourier}
\usepackage{theorem}
\usepackage{epsfig}
\usepackage{enumerate}
\usepackage{color,xcolor}
\usepackage{xspace,needspace,xparse,easyReview}
\usepackage[active]{srcltx}
\usepackage[colorlinks=true]{hyperref}
\usepackage{fancyhdr}
\usepackage{xargs}
\usepackage[toc]{appendix}

\newcommandx{\ca}[2][1=]{\todo[inline,author={ca},
	linecolor=green,backgroundcolor=green!15,bordercolor=green,#1]{#2}}
\newcommandx{\canot}[2][1=]{\todo[author={ca},
	linecolor=green,backgroundcolor=green!15,bordercolor=green,#1]{#2}}
\colorlet{darkblue}{blue!50!black}
\hypersetup{
	colorlinks,%
	citecolor=darkblue,%
	filecolor=red,%
	linkcolor=darkblue,%
	urlcolor=darkblue,%
	pdfnewwindow=true,%
	pdfstartview={FitH}
}

\numberwithin{equation}{section}

\def\init{\setcounter{equation}{0}}

\newcounter{smallarabics}

\newcounter{smallroman}

\newcommand{\ben}{\begin{enumerate}[{\rm (1)}]}
\newcommand{\een}{\end{enumerate}}

\newtheorem{theorem}{Theorem}[section]
\newtheorem{proposition}[theorem]{Proposition}
\newtheorem{lemma}[theorem]{Lemma}

\newtheorem{corollary}[theorem]{Corollary}

\usepackage[normalem]{ulem}

\NewDocumentCommand{\ASS}{mm}{\expandafter\newcommand\csname #1\endcsname{{\hyperref[#1]{\bf (#2)}}}}
\NewDocumentCommand{\preASS}{mm}{\expandafter\newcommand\csname pre#1\endcsname{{\hyperref[#1]{\bf (#2)}}}}

\AtBeginEnvironment{theorem}{\Needspace{5\baselineskip}}
\AtBeginEnvironment{proposition}{\Needspace{3\baselineskip}}
\AtBeginEnvironment{definition}{\Needspace{5\baselineskip}}
\AtBeginEnvironment{corollary}{\Needspace{5\baselineskip}}
\AtBeginEnvironment{lemma}{\Needspace{5\baselineskip}}
\AtBeginEnvironment{quote}{\Needspace{5\baselineskip}}

 \def\cH{{\mathcal H}}

\def\i{\mathrm{i}}

\def\d{\mathrm{d}}
\def\sing{\mathrm{sing}}

\def\bep{\begin{proposition}}
\def\eep{\end{proposition}}
\def\bet{\begin{theorem}}
\def\eet{\end{theorem}}
\def\bel{\begin{lemma}}
\def\eel{\end{lemma}}

\def\Im{\mathop{\mathrm{Im}}}

\newcommand{\one}{\mathbbm{1}}

\def\proof{\noindent {\bf Proof.}\ \ }
\def\qed{\hfill $\Box$\medip}

\def\textsl{{}}

\def\c0inf{C_0^\infty}
\def\proof{\noindent {\bf Proof.}\ \ }

\def\i{\mathrm{i}}
\newcommand{\beq}{\begin{equation}}
\newcommand{\eeq}{\end{equation}}
\newcommand{\bear}[1]{\begin{array}{#1}}
\newcommand{\ear}{\end{array}}

\renewcommand{\d}{\mathrm{d}}

\def\qed{$\Box$\medskip}

\def\bar{\overline}

\def\supp{{\rm supp}}
\def\ac{{\rm ac}}

\def\rr{{\mathbb R}}
\def\zz{{\mathbb Z}}
\def\cc{{\mathbb C}}

\def\P{{\mathbb P}}
\def\E{{\mathbb E}}
\def\cC{{\mathcal C}}

\def\bec{\begin{corollary}}
\def\eec{\end{corollary}}

\begin{document}
\def\today{}
\title{Extended States on the Bethe Lattice Revisited}
\author{Vojkan  Jak\v{s}i\'c$^{1}$ \,\,and \,\,Yoram  Last$^{2}$
\\ \\
$^1$Dipartimento di Matematica\\
Politecnico di Milano, 
Piazza Leonardo da Vinci 32\\
20133 Milano, Italy
\\ \\
$^2$Einstein Institute of Mathematics\\
The Hebrew University of Jerusalem\\
Edmond J. Safra Campus, Givat Ram\\
9190401 Jerusalem, Israel
}
\date{}
\maketitle
\thispagestyle{empty}

{\small
{\bf Abstract.} We give a short proof of the classical weak-disorder result that the Anderson model on
the Bethe lattice has purely absolutely continuous spectrum on compact intervals in the interior of the free spectrum.
This note continues~\cite{JLWextended}, where we gave a minimal proof of the existence of a nontrivial absolutely
continuous component using the cyclicity criterion of~\cite{JLsimple}. There, Hellinger overlap yields the required non-cyclicity.
Here, a stability argument for the forward Green function, followed by the zero--one law for the tree recursion,
yields positivity of its boundary imaginary part. This positivity already gives absolutely continuous spectrum throughout the interval;
the structural result of~\cite{JLstructure} excludes singular spectrum and yields  purity.
We also observe that the same structural argument shortens the final passage from boundary positivity to purity in the recent OpenAI preprint on the weak-disorder Anderson model on $\zz^d$, $d\geq3$~\cite{OpenAIAnderson2026}.
}

\section{Introduction}
We consider the Anderson model on the Bethe lattice at weak disorder and give
a new proof of the well-known result that, on every compact interval in the
interior of the free spectrum, the spectrum is purely absolutely continuous
for sufficiently small disorder.  This note continues~\cite{JLWextended},
where the cyclicity criterion of~\cite{JLsimple} was used to deduce the
existence of a nontrivial absolutely continuous component from Hellinger
overlap.  Here boundary positivity already gives absolutely continuous
spectrum on the interval; the structural result of~\cite{JLstructure} is
used   to exclude singular spectrum and  obtain purity.

Two structural results enter the two notes, but in different ways.  The first is
\cite[Theorem~1.2 and Remark~2]{JLsimple},
\beq
\cH_{\omega,\sing}\subset\cC_\omega(\delta_x)
\label{JL-orth-intro}
\eeq
for every site $x$ and for $\P$-a.e.\ $\omega$.  In~\cite{JLWextended},
Hellinger overlap gives non-cyclicity, and (\ref{JL-orth-intro}) turns this
into the existence of a nontrivial absolutely continuous component; the same
viewpoint also leads there to an integrated resolvent criterion for the
Anderson model on $\zz^d$ which implies non-cyclicity.  The second result
is~\cite{JLstructure}: the absolutely continuous part has a nonrandom
essential support, and the singular part vanishes almost surely on each fixed
deterministic representative of this support.  In the present note boundary
positivity already gives absolutely continuous spectrum, and
\cite{JLstructure} is used only to eliminate singular spectrum and obtain
purity.  Unlike~\cite{JLWextended}, the present argument does not yield a new
criterion for the Anderson model on $\zz^d$.

The analytic input in the present note is stronger but still minimal.  On a
compact interval in the interior of the free spectrum we prove that, as the
disorder and the imaginary part of the energy tend to zero, the forward Green
function converges in probability to the unique fixed point in $\cc_+$ of the
free recursion.  The proof uses the quantity
\[
d(z,w)=\frac{|z-w|^2}{\Im z\,\Im w},
\qquad z,w\in\cc_+,
\]
and consists of a variance estimate, one strict-convexity identity, and the
free fixed-point equation.  This hyperbolic-geometry mechanism was introduced
in this context by Froese--Hasler--Spitzer~\cite{FHS} and developed further
in~\cite{KLW,DS}.  The stability argument used here is a simplified
sequential form of the elliptic argument of Drogin--Smart~\cite[Section~4]{DS};
the precise correspondence is given in Section~\ref{sec-stability}.

To pass from this stability statement to almost-sure boundary positivity, we
use the zero--one law of~\cite[Lemma~4.1]{AW}, which follows from the tree
recursion.  For the rooted-tree operator, set
\[
q(E)=
\P\left\{
\Im\langle\delta_0,(H_\omega-E-\i0)^{-1}\delta_0\rangle=0
\right\}.
\]
Then, for almost every $E$, $q(E)=q(E)^K$.  Proposition~\ref{proposition-stability}
excludes the value one at sufficiently weak disorder, and hence
\[
\Im\langle\delta_0,(H_\omega-E-\i0)^{-1}\delta_0\rangle>0
\]
almost surely for Lebesgue-a.e.\ $E$ in every compact interval in the
interior of the free spectrum.  The Schur complement gives the same
positivity for the resolvent matrix element at the vertex $0$ of the full-tree
operator.  This already gives absolutely continuous spectrum on the
interval; Corollaries~1.1.1 and~1.1.3 of~\cite{JLstructure} eliminate the
singular spectrum there and give purity.

The comparison with~\cite{JLWextended} is particularly transparent when
written directly in terms of resolvent matrix elements.  In both conditions
below, $H_\omega$ denotes the Anderson model on the rooted tree and
$\delta_0$ its root vector. \footnote{Decoupling one vertex of the Bethe lattice
leaves $K+1$ copies of the rooted tree.  In the present setting, both the
existence of a nontrivial absolutely continuous component and pure absolute
continuity on $I$ are equivalent for the rooted and full-tree operators; the
passage between them follows from the resolvent recursion and the Schur
complement.} In~\cite{JLWextended}, the estimate proved on a
compact interval $I$ in the interior of the free spectrum is
\beq
\liminf_{\eta\downarrow0}
\int_I
\left(
\E\sqrt{
\Im\langle\delta_0,(H_\omega-E-\i\eta)^{-1}\delta_0\rangle
}
\right)^2
\,\d E>0.
\label{marseille-1}
\eeq
By the Hellinger-overlap criterion of~\cite{JLWextended},
\eqref{marseille-1} is equivalent to the existence of a nontrivial
absolutely continuous component in $I$.

The corresponding condition in the present note is
\beq
\Im\langle\delta_0,(H_\omega-E-\i0)^{-1}\delta_0\rangle>0
\label{marseille-2}
\eeq
for Lebesgue-a.e.\ $E\in I$, almost surely.\footnote{By Fatou's lemma,
\eqref{marseille-2} implies \eqref{marseille-1}.}
By~\cite{JLstructure}, \eqref{marseille-2} is sufficient for purely
absolutely continuous spectrum on $I$, and, if the single-site distribution
is equivalent to Lebesgue measure, it is also necessary.\footnote{Fix all
random potentials except the one at the root.  The claim then follows from
Theorem~55 of~\cite{JaksicSpectral} and the rank-one resolvent formula.}

The theorem proved here is well known, and much stronger results are available. 
We distinguish between results establishing purity and those establishing persistence of the absolutely continuous component without proving purity. Purely absolutely continuous spectrum on compact intervals in the interior of the free spectrum at sufficiently small disorder was proved in~\cite{Klein94,Klein98}; \cite{Klein98} also shows the integrated density of states is continuously differentiable on these intervals. Klein’s hypotheses allow unbounded and singular single-site distributions. A geometric proof of pure absolutely continuous spectrum at weak disorder for the degree-three tree was given in~\cite{FHS}.
For a larger class of trees of finite cone type, \cite{KLW} proves purely absolutely continuous spectrum and uniform \(L^p\) convergence of the Green function at weak disorder on compact energy sets avoiding a finite exceptional set. The work~\cite{ASW} establishes persistence and \(L^1\)-stability of the absolutely continuous spectral density in a setting that includes radially periodic backgrounds and weak correlations between different branches, but does not establish purity. 
The resonant-delocalization results of~\cite{AW} establish the presence of absolutely continuous spectrum beyond the free spectrum and, under their distributional hypotheses, purity near the spectral edges for bounded potentials. For unbounded single-site distributions with support \(\mathbb R\), the weak-disorder result of~\cite{AW} establishes absolutely continuous spectrum on compact intervals in \((-K-1,K+1)\), but not purity.
More recently, under fourth-moment and density regularity assumptions, \cite{DS} proved purely absolutely continuous spectrum in \(|E|<K+1-\varepsilon\) at sufficiently small disorder. In particular, for bounded single-site potentials satisfying the regularity assumption of~\cite{DS} and sufficiently small disorder, the entire spectrum is purely absolutely continuous.

We do not improve any of these results.  In the regimes where they are
directly comparable, the available results are substantially stronger than
the theorem proved here.  Instead, we isolate a direct minimal route to pure
absolutely continuous spectrum on compact intervals in the interior of the
free spectrum.  The model-specific analysis is reduced to the stability
statement and the tree recursion; once boundary positivity is obtained,
\cite{JLstructure} is needed only for the final passage to purity.

The conceptual difference between the two notes requires a comment.  In
\cite{JLWextended}, the structural result changes the problem one has to
solve: instead of proving absolute continuity directly, it is enough to prove
non-cyclicity, which can be detected by Hellinger overlap.  This viewpoint
also leads there to an integrated resolvent condition for the Anderson model
on $\zz^d$ which implies non-cyclicity and hence the existence of absolutely
continuous spectrum.  In the present note there is no comparable structural
reformulation of the existence problem.  Boundary positivity already gives
absolutely continuous spectrum by standard spectral theory; the role of
\cite{JLstructure} is to exclude singular spectrum and upgrade absolute
continuity on the interval to purity.

Finally,  we emphasize that the analytic arguments of the two notes are quite different. In~\cite{JLWextended}, the contradiction hypothesis drives the imaginary part of the forward Green function to zero, so the analysis takes place near the boundary of $\cc_+$. One must identify the relevant scale, pass to limiting laws, and use the recursion to exclude the possible boundary limits. In the present argument, by contrast, the relevant stability mechanism takes place around the free fixed point,  which has imaginary part uniformly bounded away from zero on compact intervals in the interior of the free spectrum.  The variance estimate and strict convexity force concentration, the fixed-point equation identifies the concentration point, and the zero--one law converts this stability into boundary positivity. Thus the stronger spectral conclusion is obtained from a more rigid and technically simpler analytic mechanism.

After this note was completed, OpenAI~\cite{OpenAIAnderson2026} released a
preprint proving purely absolutely continuous spectrum at weak disorder for
the Anderson model on $\zz^d$, $d\geq3$, on an open energy interval.
The proof first shows that the boundary imaginary part is positive with
positive probability and then derives purity by propagating this positivity
and using conditional spectral averaging.  Appendix~\ref{app:purity-consequence}
shows that Corollaries~1.1.1 and~1.1.3 of~\cite{JLstructure}, used here for
the Bethe lattice, already give purity directly from that boundary statement.

\paragraph*{Acknowledgments.}
This work was partly funded by the CY Initiative grant
\emph{Investissements d'Avenir}, grant number ANR-16-IDEX-0008.
V.J. acknowledges the support of the MUR grant
``Dipartimento di Eccellenza 2023--2027'' of the Dipartimento di Matematica,
Politecnico di Milano.
We also acknowledge the support of the ANR project DYNACQ,
grant number ANR-24-CE40-5714.

V.J. wishes to thank Simone Warzel for discussions during the Les Houches summer school in August 2026 that ultimately led to this note.
\section{Setup and main result}
\label{sec-setup}
Let $K\geq2$.  We denote by $\mathbb T$ the regular rooted tree of branching
number $K$, with root $0$, and by $\mathbb B$ the regular tree of degree
$K+1$.  Removing one vertex of $\mathbb B$ leaves $K+1$ copies of
$\mathbb T$.

The adjacency operators on $\mathbb T$ and $\mathbb B$ will both be denoted
by $A$, the underlying graph being clear from the context.  Let $\{V_x\}$ be
independent, identically distributed real random variables with common
absolutely continuous probability distribution $\nu$ supported in
$[-v_*,v_*]$.
We consider
\[
H^+_{\lambda,\omega}=A+\lambda V_\omega
\quad\hbox{on }\ell^2(\mathbb T),
\qquad
H_{\lambda,\omega}=A+\lambda V_\omega
\quad\hbox{on }\ell^2(\mathbb B).
\]
The spectra of these operators are $\P$-almost surely constant.  In the present
normalization (see~\cite{AW}),
\beq
\sigma(H^+_{\lambda,\omega})
=
\sigma(H_{\lambda,\omega})
=
[-2\sqrt K,2\sqrt K]+\lambda\,\supp\nu
\label{almost-sure-spectrum}
\eeq
for $\P$-a.e.\ $\omega$.  
Thus the compact intervals considered below remain inside the almost-sure
spectrum for sufficiently small disorder.
At the root of $\mathbb T$, set
\beq
\Gamma_\lambda(z)
=
\langle\delta_0,(H^+_{\lambda,\omega}-z)^{-1}\delta_0\rangle,
\qquad z\in\cc_+.
\label{forward-green}
\eeq
Let $\Gamma_{\lambda,1}(z),\ldots,\Gamma_{\lambda,K}(z)$ be independent
copies of $\Gamma_\lambda(z)$, and let $V_\omega(0)$ be independent of them.  Then
\beq
\Gamma_{\lambda,0}(z)
=
\frac1{\lambda V_\omega(0)-z-\sum_{j=1}^K\Gamma_{\lambda,j}(z)}
\label{cavity-recursion}
\eeq
has the same distribution as $\Gamma_\lambda(z)$.

For $\lambda=0$, the forward Green function is the deterministic solution
$g(z)\in\cc_+$ of
\beq
g(z)=\frac1{-z-Kg(z)}.
\label{free-fixed-point}
\eeq
In particular,
\beq
g(E+\i0)
=
\frac{-E+\i\sqrt{4K-E^2}}{2K},
\qquad |E|<2\sqrt K,
\label{free-boundary}
\eeq
and
\beq
|g(E+\i0)|=\frac1{\sqrt K}.
\label{free-modulus}
\eeq

We denote by  $\cH_{\omega,\sing}$  the singular spectral subspace of
$H_{\lambda,\omega}$.

\bet
Let $I\Subset(-2\sqrt K,2\sqrt K)$ be a nonempty compact interval.  There is
$\lambda_I>0$ such that, for every $|\lambda|<\lambda_I$, $I$ is contained in the essential support of
the absolutely continuous spectrum and 
\beq
\one_I(H_{\lambda,\omega})\cH_{\omega,\sing}=\{0\}
\label{pure-ac-conclusion}
\eeq
for $\P$-a.e.\ $\omega$.  
\label{theorem-pure-ac}
\eet

For $\lambda=0$ this is immediate, and below we suppose $\lambda\ne0$.

\subsection{Three elementary facts}
For $z,w\in\cc_+$ set
\beq
d(z,w)=\frac{|z-w|^2}{\Im z\,\Im w}.
\label{hyperbolic-d}
\eeq
The semimetric $d$ and its contraction properties play a role 
in the geometric approach to tree Green functions; see, for example,
\cite{FHS,KLW,DS}.  We collect the results we need  and include the
short proofs for the reader's convenience.

\bel
Let $a>0$.  The following statements hold.
\ben
\item If $\phi$ is a real M\"obius automorphism of $\cc_+$, then
\[
d(\phi(z),\phi(w))=d(z,w).
\]
\item If $\Im w\geq a$ and $u\in\cc$ satisfies $\Im u\geq0$ and $|u|\leq1$,
then
\beq
d(z+u,w)\leq(1+C_a|u|)d(z,w)+C_a|u|.
\label{small-shift}
\eeq
\item If $z_1,\ldots,z_K,w\in\cc_+$ and
$\bar z=K^{-1}\sum_jz_j$, then
\beq
d(\bar z,w)\leq\frac1K\sum_{j=1}^Kd(z_j,w)
\label{ordinary-convexity}
\eeq
and
\beq
d(\bar z,w)
\leq
\left(
1-
\frac{\sum_{j<k}|z_j-z_k|^2}
{K\sum_{k=1}^K|z_k-w|^2}
\right)
\max_{1\leq k\leq K}d(z_k,w),
\label{strict-convexity}
\eeq
where the fraction is understood to be zero if all $z_k=w$.
\een
\label{lemma-hyperbolic-elementary}
\eel

\proof
The first assertion follows by direct calculation for translations by real
numbers, positive dilations, and $z\mapsto-1/z$, which generate the real
M\"obius automorphisms of $\cc_+$.

For the second, first note that
\beq
\frac{\Im z}{\Im w}+\frac{\Im w}{\Im z}
\leq2+d(z,w).
\label{imaginary-ratio}
\eeq
Indeed, the right-hand side dominates
$(\Im z-\Im w)^2/(\Im z\Im w)+2$.
Since $\Im(z+u)\geq\Im z$,
\[
\sqrt{d(z+u,w)}
\leq
\sqrt{d(z,w)}+
\frac{|u|}{\sqrt{\Im z\,\Im w}}.
\]
Using (\ref{imaginary-ratio}) and $\Im w\geq a$, and then squaring, gives
(\ref{small-shift}).

For (\ref{ordinary-convexity}), weighted Cauchy--Schwarz gives
\[
\frac{|\sum_j(z_j-w)|^2}{\sum_j\Im z_j}
\leq
\sum_j\frac{|z_j-w|^2}{\Im z_j},
\]
and division by $K\Im w$ yields the assertion.
Finally, set
\[
D=\max_{1\leq k\leq K}d(z_k,w).
\]
If $D=0$, then all $z_k=w$ and there is nothing to prove.  Otherwise,
\[
|z_k-w|^2\leq D\,\Im z_k\,\Im w,
\]
and therefore
\[
D\,\Im w\sum_k\Im z_k
\geq
\sum_k|z_k-w|^2.
\]
Since
\[
d(\bar z,w)
=
\frac{|\sum_k(z_k-w)|^2}
{K\Im w\sum_k\Im z_k},
\]
we obtain
\[
d(\bar z,w)
\leq
\frac{|\sum_k(z_k-w)|^2}
{K\sum_k|z_k-w|^2}D.
\]
The elementary identity
\[
|\sum_k(z_k-w)|^2
=
K\sum_k|z_k-w|^2-
\sum_{j<k}|z_j-z_k|^2
\]
gives (\ref{strict-convexity}).
\qed

\subsection{Stability}
\label{sec-stability}
We now prove that, as the disorder and the imaginary part of the energy vanish, the forward Green function converges in probability to the free fixed point. Combined with the zero--one law of Section~\ref{sec-zero-one}, this will imply positivity of the boundary imaginary part on compact intervals in the interior of the free spectrum.

The proposition below is a sequential form of the elliptic stability result proved in~\cite[Section~4]{DS}; its variance, concentration, and fixed-point steps correspond respectively to Lemmas~4.2, 4.4, and~4.6 there. For our purposes only this weaker statement is needed, and in this form the proof is short and self-contained.
\bep
Let $I\Subset(-2\sqrt K,2\sqrt K)$.  Suppose $\lambda_n\rightarrow0$, $E_n\in I$, $\eta_n\downarrow0$.
Then
\beq
\Gamma_{\lambda_n}(E_n+\i\eta_n)
-
g(E_n+\i0)
\rightarrow0
\quad\hbox{in probability}.
\label{stability}
\eeq
\label{proposition-stability}
\eep


For the proof write $h_n=-\lambda_nV_\omega(0)$ and $w_n=g(E_n+\i0)$.  Since the
potential is bounded,
\beq
|h_n|\leq C|\lambda_n|
\quad\hbox{almost surely},
\label{hn-small}
\eeq
while, since $E_n\in I$,
\beq
\Im w_n\geq c_I>0,
\qquad
|w_n|=K^{-1/2}.
\label{wn-elliptic}
\eeq
Let $X_{n,1},\ldots,X_{n,K}$ be independent copies of
$X_n=\Gamma_{\lambda_n}(E_n+\i\eta_n)$, independent also of $h_n$, and
\[
X_{n,0}
=
-\frac1{X_{n,1}+\cdots+X_{n,K}+h_n+E_n+\i\eta_n}.
\]
By \eqref{cavity-recursion} and self-similarity, \(X_{n,0}\) has the same law as \(X_n\). We set
\[
D_{n,j}=d(X_{n,j},w_n),
\qquad
M_n=\E D_{n,0}^2,
\qquad
\bar d_n=\E D_{n,0}.
\]
For every fixed $n$, $M_n<\infty$.  Indeed, $|X_{n,j}|\leq\eta_n^{-1}$,
while the recursion gives
\[
\Im X_{n,0}
\geq
\frac{\eta_n}{(K\eta_n^{-1}+C)^2}.
\]
Finally, let
\[
A_n=\frac1K\sum_{j=1}^KX_{n,j},
\qquad
u_n=\frac{h_n+\i\eta_n}{K},
\qquad
\phi_n(z)=-\frac1{Kz+E_n}.
\]
The map $\phi_n$ is a real M\"obius automorphism of $\cc_+$ that fixes $w_n$,
and $X_{n,0}=\phi_n(A_n+u_n)$.

Set
\[
\varepsilon_n=|\lambda_n|+\eta_n.
\]
\bel For some $C>0$,
\beq
\operatorname{Var}(D_{n,0})
\leq
C\varepsilon_n(1+M_n).
\label{variance-bound}
\eeq
\label{lemma-stability-variance}
\eel
\proof
By Lemma~\ref{lemma-hyperbolic-elementary}, ordinary convexity, and
(\ref{hn-small})--(\ref{wn-elliptic}), there exists  $C>0$ such that 
\beq
D_{n,0}
\leq
(1+C\varepsilon_n)\frac1K\sum_{j=1}^KD_{n,j}
+C\varepsilon_n.
\label{basic-D-inequality}
\eeq
If $Z_n=K^{-1}\sum_{j=1}^KD_{n,j}$, independence gives
\[
\E Z_n^2
=
\bar d_n^2+\frac1K\operatorname{Var}(D_{n,0})
=
M_n-\frac{K-1}{K}\operatorname{Var}(D_{n,0}).
\]
Squaring (\ref{basic-D-inequality}), taking expectations, and absorbing the
mixed term gives
\[
M_n
\leq
(1+C\varepsilon_n)
\left(
M_n-\frac{K-1}{K}\operatorname{Var}(D_{n,0})
\right)
+C\varepsilon_n.
\]
Since $\varepsilon_n\to0$, this implies (\ref{variance-bound}).
\qed


\bel
Suppose $M_n\geq c>0$.  Then
\beq
\bar d_n\geq c'>0,
\qquad
\frac{D_{n,j}}{\bar d_n}\rightarrow1
\quad\hbox{in probability},
\qquad j=0,1,\ldots,K.
\label{D-concentration}
\eeq
Moreover, there is a deterministic sequence $\delta_n\downarrow0$ such that,
for two independent copies $X_n',X_n''$ of $X_n$,
\beq
\P\{|X_n'-X_n''|>\delta_n|X_n''-w_n|\}\rightarrow0.
\label{pair-collapse}
\eeq
\label{lemma-stability-collapse}
\eel

\proof
By Lemma~\ref{lemma-stability-variance},
$\operatorname{Var}(D_{n,0})/M_n\rightarrow0$.  Since
$M_n=\bar d_n^2+\operatorname{Var}(D_{n,0})$, we have
$\bar d_n^2/M_n\rightarrow1$, and (\ref{D-concentration}) follows.

Set
\[
R_n=
\frac{\sum_{j<k}|X_{n,j}-X_{n,k}|^2}
{K\sum_{k=1}^K|X_{n,k}-w_n|^2},
\]
with $R_n=0$ when the denominator vanishes.  The strict-convexity estimate
gives
\[
D_{n,0}
\leq
(1+C\varepsilon_n)(1-R_n)
\max_{1\leq j\leq K}D_{n,j}
+C\varepsilon_n.
\]
Dividing by $\bar d_n$ and using (\ref{D-concentration}) yields
\beq
R_n\rightarrow0
\quad\hbox{in probability}.
\label{R-to-zero}
\eeq
Let $S_n=\sum_{j=1}^K|X_{n,j}-w_n|^2$.  The identities
\[
\sum_{j<k}|X_{n,j}-X_{n,k}|^2
=
K\sum_{j=1}^K|X_{n,j}-A_n|^2,
\qquad
|A_n-w_n|^2=\frac{1-R_n}{K}S_n
\]
show that, whenever $R_n\leq1/(4K)$,
\[
|X_{n,1}-w_n|\geq c_KS_n^{1/2},
\qquad
|X_{n,1}-X_{n,2}|\leq2R_n^{1/2}S_n^{1/2}.
\]
It follows from (\ref{R-to-zero}) that
\[
\frac{|X_{n,1}-X_{n,2}|}{|X_{n,1}-w_n|}
\rightarrow0
\quad\hbox{in probability}.
\]
Interchanging the two independent copies gives the same statement with
$|X_{n,2}-w_n|$ in the denominator, and (\ref{pair-collapse}) follows for a
suitable deterministic sequence $\delta_n\downarrow0$.
\qed

\bel
Suppose that $\delta_n\downarrow0$ satisfies (\ref{pair-collapse}).  Then
\beq
X_{n,0}-w_n\rightarrow0
\quad\hbox{in probability}.
\label{fixed-point-identification}
\eeq
\label{lemma-stability-fixed-point}
\eel

\proof
Let $\widetilde X_n$ be a copy of $X_n$, independent of
$h_n,X_{n,0},\ldots,X_{n,K}$.  By (\ref{pair-collapse}) and the union bound,
\beq
\P\left\{
\max_{0\leq j\leq K}|X_{n,j}-\widetilde X_n|
\leq\delta_n|\widetilde X_n-w_n|
\right\}\rightarrow1.
\label{center-event}
\eeq
On this event, $\phi_n^{-1}(X_{n,0})=A_n+u_n$ gives
\beq
|X_{n,0}-\phi_n^{-1}(X_{n,0})|
\leq
2\delta_n|\widetilde X_n-w_n|+C(|\lambda_n|+\eta_n).
\label{approx-fixed}
\eeq
The two fixed points of $\phi_n$ are $w_n$ and $\overline w_n$, and
\[
x-\phi_n^{-1}(x)
=
\frac{(x-w_n)(x-\overline w_n)}{x}.
\]
For $x\in\cc_+$, (\ref{wn-elliptic}) gives
\[
\frac{|x|}{|x-\overline w_n|}
\leq1+\frac{|w_n|}{\Im w_n}
\leq B_I,
\]
and hence $|x-w_n|\leq B_I|x-\phi_n^{-1}(x)|$.  On the event in
(\ref{center-event}),
$|\widetilde X_n-w_n|\leq|X_{n,0}-w_n|/(1-\delta_n)$.  Combining this with
(\ref{approx-fixed}) gives
\[
|X_{n,0}-w_n|
\leq
\frac{2B_I\delta_n}{1-\delta_n}|X_{n,0}-w_n|
+C_I(|\lambda_n|+\eta_n).
\]
For  sufficiently large $n$ the first term on the right can be absorbed
into the left.  This proves (\ref{fixed-point-identification}).
\qed

\noindent {\bf Proof of Proposition~\ref{proposition-stability}.}\
Suppose $M_n$ does not tend to zero.  Passing to a subsequence, we may assume
$M_n\geq c>0$.  Lemma~\ref{lemma-stability-collapse} gives
(\ref{D-concentration}) and (\ref{pair-collapse}), while
Lemma~\ref{lemma-stability-fixed-point} gives $X_{n,0}-w_n\rightarrow0$ in
probability.  Since $\Im w_n\geq c_I$, it follows that $D_{n,0}\rightarrow0$
in probability, contradicting (\ref{D-concentration}) and the lower bound on
$\bar d_n$.  Hence $M_n\rightarrow0$.

It follows that $D_{n,0}\rightarrow0$ in probability.  The imaginary-part
ratio estimate (\ref{imaginary-ratio}) gives
\[
\frac{\Im X_{n,0}}{\Im w_n}\rightarrow1
\quad\hbox{in probability},
\]
and the definition of $d$ then yields $|X_{n,0}-w_n|\rightarrow0$ in
probability.  Since $X_{n,0}$ has the same law as $X_n$, this proves
(\ref{stability}).
\qed


\section{The boundary zero--one law}
\label{sec-zero-one}
\init
For each realization, $\Gamma_\lambda$ is a Herglotz function.  Hence its
boundary value exists and is finite and nonzero for Lebesgue-a.e.\ energy; see, for example,
\cite[Theorem~35(1)]{JaksicSpectral}.   By Fubini, for
Lebesgue-a.e.\ $E$, $\Gamma_\lambda(E+\i0)$ exists and is finite and nonzero
almost surely. For such $E$, set
\[
Y_\lambda(E)=\Im\Gamma_\lambda(E+\i0),
\qquad
q_\lambda(E)=\P\{Y_\lambda(E)=0\}.
\]
The following zero--one law is Lemma~4.1 of~\cite{AW}; we include its short proof.
\bel
For Lebesgue-a.e.\ $E$,
\beq
q_\lambda(E)\in\{0,1\}.
\label{zero-one-values}
\eeq
\label{lemma-zero-one}
\eel

\proof
Consider independent copies $\Gamma_{\lambda,1},\ldots,\Gamma_{\lambda,K}$ of
$\Gamma_\lambda$ and define $\Gamma_{\lambda,0}$ by
(\ref{cavity-recursion}). For almost every $E$, all the
boundary values $\Gamma_{\lambda,j}(E+\i0)$
exist and are finite and nonzero almost surely.  Set $Y_{\lambda,j}(E)=\Im\Gamma_{\lambda,j}(E+\i0)$. 
Passing to the boundary in the imaginary
part of (\ref{cavity-recursion}) gives
\beq
Y_{\lambda,0}(E)
=
|\Gamma_{\lambda,0}(E+\i0)|^2
\sum_{j=1}^KY_{\lambda,j}(E).
\label{boundary-imaginary-recursion}
\eeq
The factor multiplying the sum is strictly positive and all
$Y_{\lambda,j}$ are nonnegative.  Thus
\[
Y_{\lambda,0}(E)=0
\quad\Longleftrightarrow\quad
Y_{\lambda,1}(E)=\cdots=Y_{\lambda,K}(E)=0.
\]
The variables on the right are independent copies of $Y_\lambda(E)$, while
$Y_{\lambda,0}(E)$ has the same law.  Hence,  $q_\lambda(E)=q_\lambda(E)^K$. 
Since $K\geq2$,  $q_\lambda(E)\in\{0,1\}$.
\qed

\bep
Let $I\Subset(-2\sqrt K,2\sqrt K)$.  For all sufficiently small
$|\lambda|$,
\beq
\P\{\Im\Gamma_\lambda(E+\i0)>0\}=1
\label{positive-boundary}
\eeq
for Lebesgue-a.e.\ $E\in I$.
\label{proposition-positive-boundary}
\eep

\proof
Suppose the assertion is false.  By Lemma \ref{lemma-zero-one}, there are
$\lambda_n\to0$ for which $\{E\in I:q_{\lambda_n}(E)=1\}$ has positive Lebesgue
measure.  Choose $E_n$ in this set such that the boundary value exists almost
surely, and set
\[
\kappa
=
\frac14\inf_{E\in I}\Im g(E+\i0)>0.
\]
Since $q_{\lambda_n}(E_n)=1$, $\Im\Gamma_{\lambda_n}(E_n+\i\eta)
\rightarrow0 $ almost surely as $\eta\downarrow0$. 
We can therefore choose $0<\eta_n<1/n$ so that
\beq
\P\{\Im\Gamma_{\lambda_n}(E_n+\i\eta_n)\geq\kappa\}<\frac14.
\label{bad-probability}
\eeq
On the other hand, Proposition~\ref{proposition-stability} gives
\[
\Gamma_{\lambda_n}(E_n+\i\eta_n)
-
g(E_n+\i0)
\rightarrow0
\]
in probability.  Since
$\Im g(E_n+\i0)\geq4\kappa$, the probability in
(\ref{bad-probability}) tends to one.  This is a contradiction.  Hence the
set on which $q_\lambda(E)=1$ has Lebesgue measure zero for all sufficiently
small $|\lambda|$, and Lemma \ref{lemma-zero-one} gives
(\ref{positive-boundary}).
\qed
\section{From boundary positivity to purity}
\label{sec-purity-principle}
\init

We now complete the proof  of Theorem~\ref{theorem-pure-ac} using Corollaries~1.1.1 and~1.1.3
of~\cite{JLstructure}.  For Schr\"odinger operators on connected graphs with
independent absolutely continuous diagonal randomness, the site vectors form a
cyclic family and the cyclic subspaces generated by any two site vectors are
non-orthogonal.  Hence every site spectral measure belongs almost surely to
the spectral measure class of the operator.  Moreover, the absolutely
continuous part has a nonrandom essential support, and the singular part
vanishes almost surely on each fixed deterministic representative of this
support.

We apply this to $H_{\lambda,\omega}$ on $\mathbb B$.  Fix a vertex $x$.
Removing $x$ leaves $K+1$ independent copies of $\mathbb T$, and the Schur
complement gives
\beq
G_{\omega,x}(z)
:=
\langle\delta_x,(H_{\lambda,\omega}-z)^{-1}\delta_x\rangle
=
\frac1{\lambda V_\omega(x)-z-\sum_{j=1}^{K+1}\Gamma_{\lambda,j}(z)}.
\label{full-tree-schur}
\eeq
By Proposition~\ref{proposition-positive-boundary} and Fubini,
\[
\Im\Gamma_{\lambda,j}(E+\i0)>0,
\qquad j=1,\ldots,K+1,
\]
for $\P$-a.e.\ $\omega$ and Lebesgue-a.e.\ $E\in I$.  Hence
\beq
\Im G_{\omega,x}(E+\i0)
=
\frac{\sum_{j=1}^{K+1}\Im\Gamma_{\lambda,j}(E+\i0)}
{\left|\lambda V_\omega(x)-E-\sum_{j=1}^{K+1}
\Gamma_{\lambda,j}(E+\i0)\right|^2}
>0.
\label{full-root-positive}
\eeq
Let $\mu_{\omega,x}$ be the spectral measure of $\delta_x$.  By the
boundary-value formula \cite[Theorem~35(2)]{JaksicSpectral},
\[
\frac{d(\mu_{\omega,x})_{\ac}}{dE}
=
\frac1\pi\Im G_{\omega,x}(E+\i0),
\]
and  the existence of absolutely continuous spectrum in $I$ already follows
from \eqref{full-root-positive}.  To obtain purity, we use
\cite{JLstructure}.  The interval $I$ is contained, modulo Lebesgue-null
sets, in the nonrandom essential support of the absolutely continuous part.
Adjoining $I$ to any fixed representative of this support gives another
representative, and Corollary~1.1.3 of~\cite{JLstructure} yields
\[
\one_I(H_{\lambda,\omega})\cH_{\omega,\sing}=\{0\}.
\]

\appendix
\section{Unbounded single-site distributions}
\label{app-generalizations}
\init

The fourth-moment extension of the stability argument is also contained, in a stronger form, in the elliptic estimates of~\cite{DS}; see in particular Lemma~2.8 and Section~4. The minor modifications needed for the argument above are given below.

The main theorem remains valid for absolutely continuous single-site distributions with finite fourth moment; in particular, the distribution may have support all of \(\rr\). The only changes in the proof concern the estimates involving the shift \(u_n\).

For $\Im w\geq a>0$, $\Im u\geq0$, the proof of
Lemma~\ref{lemma-hyperbolic-elementary} gives, without the restriction
$|u|\leq1$,
\beq
d(z+u,w)
\leq
\{1+C_a(|u|+|u|^2)\}d(z,w)
+C_a(|u|+|u|^2).
\label{unbounded-shift}
\eeq
Indeed,
\[
\sqrt{d(z+u,w)}
\leq
\sqrt{d(z,w)}+
\frac{|u|}{\sqrt{\Im z\,\Im w}},
\]
and (\ref{imaginary-ratio}) gives
\[
\frac1{\sqrt{\Im z\,\Im w}}
\leq
C_a\sqrt{2+d(z,w)}.
\]
Squaring yields (\ref{unbounded-shift}).

Assume that
\beq
\E |V_\omega(0)|^4<\infty.
\label{fourth-moment}
\eeq
With the notation used in the proof of
Proposition~\ref{proposition-stability}, set
\[
a_n=C_I(|u_n|+|u_n|^2),
\qquad
\varepsilon_n=\E(a_n+a_n^2).
\]
Then $\varepsilon_n\to0$.  Moreover, for each fixed $n$, $M_n<\infty$: the
recursion and $|X_{n,j}|\leq\eta_n^{-1}$ give
\[
(\Im X_{n,0})^{-1}
\leq
\frac{(K\eta_n^{-1}+|h_n|+C_I)^2}{\eta_n},
\]
so that (\ref{fourth-moment}) controls the second moment of
$d(X_{n,0},w_n)$.

Using (\ref{unbounded-shift}) in place of (\ref{small-shift}), and the
independence of $u_n$ from the forward variables, the calculation leading to
(\ref{variance-bound}) gives
\beq
\operatorname{Var}(D_{n,0})
\leq
C\varepsilon_n(1+M_n).
\label{unbounded-variance}
\eeq
The strict-convexity argument is unchanged, since $a_n\to0$ in probability. 
At the final fixed-point step, (\ref{approx-fixed}) is replaced by
\[
|X_{n,0}-\phi_n^{-1}(X_{n,0})|
\leq
2\delta_n|\widetilde X_n-w_n|+|u_n|.
\]
On the event in (\ref{center-event}), the argument of
Lemma~\ref{lemma-stability-fixed-point}  gives
\[
|X_{n,0}-w_n|
\leq
\frac{2B_I\delta_n}{1-\delta_n}|X_{n,0}-w_n|
+B_I|u_n|.
\]
Since $\delta_n\to0$, the first term on the right can be absorbed into the
left for sufficiently large $n$.  Hence
\[
|X_{n,0}-w_n|\leq C_I|u_n|.
\]
Since $u_n\to0$ in probability,
\[
\Gamma_{\lambda_n}(E_n+\i\eta_n)-g(E_n+\i0)
\rightarrow0
\quad\hbox{in probability}.
\]

\section{From boundary positivity to pure absolute continuity on \texorpdfstring{$\zz^d$}{Z-d}}
\label{app:purity-consequence}
\init
This appendix was written on October~7, 2026, after the completion of the
present note and one day after OpenAI publicly released the preprint
\cite{OpenAIAnderson2026},  claiming to prove purely absolutely continuous spectrum at
weak disorder for the Anderson model on $\zz^d$, $d\geq3$, on an open energy
interval.  Let $I_d$ denote the interval in that result.  For Lebesgue-a.e.\
$E\in I_d$, \cite[Eq.~(177)]{OpenAIAnderson2026} gives
\beq
\P\left\{
\Im\langle\delta_0,(H_{\lambda,\omega}-E-\i0)^{-1}\delta_0\rangle>0
\right\}>0.
\label{openai-boundary}
\eeq
The subsequent argument propagates this positivity and uses conditional
spectral averaging to exclude singular spectrum.

The same purity mechanism used in Section~\ref{sec-purity-principle}, based
on~\cite[Corollaries~1.1.1 and~1.1.3]{JLstructure}, applies directly at this
point.  Let $S$ be a fixed deterministic representative of the nonrandom
essential support of the absolutely continuous part.  Corollary~1.1.1 of
~\cite{JLstructure} and \cite[Theorem~35(2)]{JaksicSpectral} imply that
\[
\P\left\{
\Im\langle\delta_0,(H_{\lambda,\omega}-E-\i0)^{-1}\delta_0\rangle>0
\right\}=0
\]
for Lebesgue-a.e.\ $E\notin S$.  Comparison with~\eqref{openai-boundary}
therefore shows that $I_d\setminus S$ is Lebesgue-null.  Thus $S\cup I_d$
is another fixed deterministic representative of the same absolutely
continuous essential support, and Corollary~1.1.3 of~\cite{JLstructure}
gives
\[
\one_{I_d}(H_{\lambda,\omega})\cH_{\omega,\sing}=\{0\}
\]
almost surely.

Thus, once Eq.~(177) of~\cite{OpenAIAnderson2026} is established, purity
follows directly from the structural argument above.  The subsequent
propagation-of-positivity and conditional spectral-averaging argument is
unnecessary for this final step.  The argument leading to
Eq.~(177), including the finite-volume analysis and the earlier use of
spectral averaging, is unchanged.

\end{document}